\documentclass[aps,pra,floatfix,superscriptaddress,twocolumn]{revtex4-2}

\usepackage{graphicx}
\usepackage[english]{babel}
\usepackage{amsmath}
\usepackage{amssymb}
\usepackage{tensor}
\usepackage{xcolor}

\newcommand{\vbr}{\hat{\mathbf{r}}}
\newcommand{\be}{\begin{eqnarray}}
\newcommand{\ee}{\end{eqnarray}}
\newcommand{\p}{\partial}

\def\ep#1{\langle #1 \rangle}

\begin{document}

\title{Two-component repulsive atomic Fermi gases in a thin spherical shell}

\author{Yan He}
\affiliation{College of Physics, Sichuan University, Chengdu, Sichuan 610064, China}
\email{heyan$_$ctp@scu.edu.cn}

\author{Chih-Chun Chien}
\affiliation{Department of Physics, University of California, Merced, CA 95343, USA.}
\email{cchien5@ucmerced.edu}

\begin{abstract}
We present possible ground-state structures of two-component atomic Fermi gases with repulsive interactions in a thin spherical shell geometry by implementing a self-consistent Hartree-Fock approximation. The system exhibits a miscible-immiscible transition from a homogeneous mixture to two-chunk phase separation as the interaction strength crosses a critical value. While the critical value is relatively insensitive to population imbalance for equal-mass mixtures, it decreases with the mass ratio when mass-imbalance is present. The interaction may be tuned by the two-body scattering length or the radius of the sphere, thereby allowing the system to cross the transition by varying different parameters. When the atoms on the sphere are rotating, three-chunk sandwich structures emerge in mass-imbalanced mixtures as a consequence of maximal angular momentum along the rotation axis. Some indications of geometric effects and possible experimental implications are also discussed.
\end{abstract}

\maketitle

\section{Introduction}
The possibilities of thin spherical shell structures for ultracold atoms by spherical bubble traps in outer space~\cite{Carollo22,Lundblad_2023} or multi-species phase separation in spherical harmonic traps~\cite{PhysRevLett.100.185301,PhysRevLett.97.060403,PhysRevLett.129.243402} introduce an interesting geometry for studying geometric effects on many-body physics (also see Refs.~\cite{TononiReview23,TONONI20241} for a review). There have been plenty of studies on bosons in a spherical-shell geometry, including Bose-Einstein condensation (BEC)~\cite{SphericalBECPRL19,SphericalSFPRL20,SphericalBECNJP20}, quantum vortices~\cite{PhysRevA.102.043305,PhysRevA.103.053306,PhysRevA.109.013301}, multi-component mixtures~\cite{PhysRevA.104.033318,PhysRevA.106.013309}, and others~\cite{PhysRevA.75.013611,PhysRevA.104.063310,PhysRevResearch.4.013122,He2023Soliton,Boegel_2023,PhysRevA.107.023319,PhysRevLett.132.026001,Arazo_2021}. In contrast, there have been relatively few studies on fermions in a similar geometry, addressing noninteracting fermions~\cite{FreeFermionShpere}, BCS-BEC crossover of Fermi superfluids~\cite{He22Sphere}, and vortex structures~\cite{He23Vortex}. 

One important difference between bosonic and fermionic superfluids is the relation between the density and order parameter. The density is the square of the order parameter in the Bogoliubov theory of bosonic BEC, but the two are independently calculated in the BCS theory of Fermi superfluids~\cite{Fetter_book,Pethick-BEC,Ueda-book}. The particular properties of fermions combined with the spherical-shell geometry results in the curvature-induced BCS-BEC crossover due to a competition between kinetic and interaction energies~\cite{He22Sphere} and counter-flowing in-gap states in higher-vorticity vortex cores~\cite{He23Vortex} without bosonic counterparts. 
Meanwhile, two-component Fermi gases with repulsive interactions do not form Cooper pairs, so there is no superfluidity in its ground state, which is different from two-component bosons that can form BEC for each component. 
Nevertheless, two-component bosons or fermions with repulsive inter-component repulsion may exhibit a miscible-immiscible transition in real space as the repulsion increases~\cite{phase-sep-exp,PhysRevLett.110.165302,Pecak16,Bellotti2017,PhysRevA.93.023612,parajuli2019mass}. The structural transition is due to a competition between the kinetic and interaction energies, which transcends spin statistics and geometry.

Nevertheless, adding confining geometries, such as a box potential, may lead to interesting structures of two-component bosons or fermions depending on the mass ratio of the mixture due to the distortion of the wavefunctions at the boundary~\cite{parajuli2019mass}. Here we consider equal-mass and mass-imbalanced two-component Fermi gases in a thin spherical shell and characterize possible ground-state structures with and without rotation of the atoms. In general, the mixture is homogeneous when the repulsion is weak but transitions to phase separation as the coupling constant exceeds a critical value. We found that the mass-imbalance induced sandwich structures of repulsive two-component Fermi gases in a box~\cite{parajuli2019mass} do not emerge on a static sphere. This contrasts the two geometries, a box and a spherical shell, since the latter is periodic in every direction and has no hard wall to distort the wavefunctions. Nevertheless, we will show that adding rotation to the atoms on the sphere will induce sandwich structures on a spherical surface because of the maximization of angular momentum along the rotation axis.

For equal-mass mixtures without rotation, the area occupied by each species in a phase-separation structure is determined by the populations of the two components. The critical interaction strength, however, is quite insensitive to population imbalance. In contrast, the lighter species in a mass-imbalanced mixture typically occupies a larger area due to its advantage of relatively higher kinetic energy. Moreover, the critical interaction strength decreases as the mass ratio increases. By adding rotation of the atoms on the sphere to the problem, we found more regimes covered by phase separation in mass-imbalanced cases due to the stabilization of sandwich structures below the critical coupling strength. Therefore, the spherical-shell geometry on its own gives rise to interesting structures of repulsive Fermi gases.

The rest of the paper is organized as follows. Sec.~\ref{Sec:Theory} outlines the mean-field theory of repulsive Fermi gases in a thin spherical shell. Population-imbalance and mass-imbalance effects are considered as well as rotation of the atoms on the sphere. An iteration method for obtaining self-consistent solutions and comparisons of ground-state energies are also explained. Sec.~\ref{Sec:Results} presents possible structures of repulsive Fermi gases on the surface of a sphere as the interaction, population imbalance, mass imbalances, and rotation are varied. A phase diagram showing where interesting structures may survive is also shown. Sec.~\ref{Sec:Discussion} discusses possible implications of our work and connections to experiments. Finally, Sec.~\ref{Sec:Conclusion} concludes our work.

\section{Theory of repulsive fermions on spherical shell}\label{Sec:Theory}
We consider two-component atomic Fermi gases with repulsive contact interactions on a spherical surface with radius $R$. The thickness of the shell is assumed to be thin enough that the radial degrees of freedom are frozen. The Hamiltonian of the system is given by
\be
H=\int_{S^2}d^2x \sqrt{\textrm{g}}\Big[\sum_a\frac{\hbar^2}{2m_a}\textrm{g}^{\mu\nu}(\p_\mu\psi_a)^\dag \p_\nu\psi_a+\tilde{g}_{12} n_1n_2\Big]. \nonumber \\
\ee
Here $m_a$ with $a=1,2$ are the masses of the two species of fermions. $n_a=\psi_a^\dag\psi_a$ are the density operators. 
We assume that $x^{\mu}$ with $\mu=1,2$ parameterize the sphere, and $\textrm{g}_{\mu\nu}$ is the metric of the sphere with $\textrm{g}=\det(\textrm{g}_{\mu\nu})$. The integration is over $S^2$, the whole surface of the sphere. If mass imbalance is present, we assume $m_1<m_2$. The coupling constant $\tilde{g}_{12}>0$ models the inter-component repulsion. 
There is no intra-component interaction between identical fermions due to the suppression of two-body $s$-wave scattering by the Pauli exclusion principle.

Following the Hartree-Fock (HF) approximation~\cite{Fetter_book}, we rewrite the interaction term by replacing one of $n_a$ by its expectation value $\ep{n_a}$.  After dropping a scalar term, we arrived at the following mean-field Hamiltonian as
\be
H_{mf}&=&\int_{S^2}d^2x \sqrt{\textrm{g}}
\Big[\sum_a\frac{\hbar^2}{2m_a}\textrm{g}^{\mu\nu}(\p_\mu\psi_a)^\dag \p_\nu\psi_a+ \nonumber \\
& & \tilde{g}_{12}\ep{n_2}\psi^\dag_1\psi_1+\tilde{g}_{12}\ep{n_1}\psi^\dag_2\psi_2\Big].
\ee
A variation of $H_{mf}$ with respect to $\psi^\dag_a$ leads to the following HF equation
\be
&&-\frac{\hbar^2}{2m_1R^2}\nabla_s^2\psi_{1,n}(\vbr)+\tilde{g}_{12}\ep{n_2}(\vbr)\psi_{1,n}(\vbr)=\tilde{E}_n^{(1)}\psi_{1,n}(\vbr), \nonumber \\
&&-\frac{\hbar^2}{2m_2R^2}\nabla_s^2\psi_{2,n}(\vbr)+\tilde{g}_{12}\ep{n_1}(\vbr)\psi_{2,n}(\vbr)=\tilde{E}_n^{(2)}\psi_{2,n}(\vbr).
\label{eq-HF}
\ee
Here $\vbr=(\theta,\phi)$. The spherical Laplacian operator is given by $\nabla_s^2\equiv-\frac{1}{\sqrt{\textrm{g}}}\p_\mu\sqrt{\textrm{g}}\textrm{g}^{\mu\nu}\p_\nu$. Explicitly,
\be
\nabla_s^2
=-\Big(\frac{1}{\sin\theta}\frac{\p}{\p\theta}\sin\theta\frac{\p}{\p\theta}+\frac{1}{\sin^2\theta}\frac{\p^2}{\p^2\phi}\Big).
\ee
We introduce the dimensionless densities $\rho_a=A\ep{n_a}$, where $A=4\pi^2R^2$ is the total area of the sphere.
After dividing both sides of Eq.~(\ref{eq-HF}) by $E_0=\frac{\hbar^2}{2m_1R^2}$, we arrived at the dimensionless version of the HF equation as
\be
&&-\nabla_s^2\psi_{1,n}(\vbr)+g_{12}\rho_2(\vbr)\psi_{1,n}(\vbr)=E_n^{(1)}\psi_{1,n}(\vbr), \nonumber \\
&&-\frac{m_1}{m_2}\nabla_s^2\psi_{2,n}(\vbr)+g_{12}\rho_1(\vbr)\psi_{2,n}(\vbr)=E_n^{(2)}\psi_{2,n}(\vbr).
\label{eq-HF1}
\ee  
Here the dimensionless coupling constant is given by $g_{12}=\tilde{g}_{12}/(E_0 A)$ and $E^{(a)}_n=\tilde{E}^{(a)}_n/E_0$ for $a=1,2$. 
In the following, we will focus on the ground state and treat the fermion operators as wave functions. The eigen-functions satisfy the orthonormal condition as
\be\label{Eq:rho}
\int_{S^2}d^2x\sqrt{\textrm{g}}\,\psi^*_{a,m}(\vbr)\psi_{b,n}(\vbr)=\delta_{ab}\delta_{mn}.
\ee
The profiles of the fermion densities are given by
\be
\ep{n_a}(\vbr)=\sum_{n=1}^{N_a}|\psi_{a,n}(\vbr)|^2.
\label{eq-num}
\ee
Here the summation is from the lowest eigen-energy state up to a fixed fermion numbers $N_a$ with $a=1,2$.

We mention that the mean-field treatment of the 2D Fermi gas on a sphere respects scale invariance~\cite{PhysRevA.55.R853}, but a consideration of the two-body scattering length has shown to introduce observable corrections~\cite{PhysRevLett.121.120401}. Similarly, two-component Fermi gases with attractive interactions associated with the two-body scattering length on a sphere described by the self-consistent Bogoliubov-de Gennes equation are shown to exhibit a size-induced BCS-BEC crossover~\cite{He22Sphere}. The relation between the coupling constant and scattering length for Fermi gases on a sphere is given by~\cite{He22Sphere} $1/\tilde{g}_{12}=\int (dl/A) (2l+1)/(\epsilon_l +|\epsilon_b|)$. Here $\epsilon_l=l(l+1)E_0$ is the dispersion of noninteracting Fermi gases on a sphere, $\epsilon_b=-\hbar^2/(2\tilde{m}a^2)$ is the two-body binding energy in 2D, $\tilde{m}$ is the reduced mass, and $a$ is the 2D two-body scattering length. For the repulsive Fermi gases considered here, we are far away from the Feshbach resonance. Thus, $a\ll R$ and $\epsilon_l\ll |\epsilon_b|$. Therefore, $1/\tilde{g}_{12}\rightarrow\Lambda\frac{\tilde{m}}{2\pi\hbar^2}(a/ R)^2$, where $\Lambda$ denotes the ultra-violet cutoff on the sphere associated with the highest angular momentum quantum number. The dimensionless coupling constant is then 
\be
g_{12}=\frac{\tilde{g}_{12}}{E_0 A}
=\frac{m_1}{\tilde{m}}\frac{1}{\Lambda} (R/a)^2,\label{eq-g12}
\ee
which increases with $R$ when the scattering length and cutoff are fixed.

Now we can expand $\psi_{1,n}(\vbr)$ and $\psi_{2,n}(\vbr)$ by spherical harmonics as
\be
&&\psi_{1,n}(\vbr)=\sum_{l,m}c_{nlm}Y_{l,m}(\theta,\phi), \nonumber \\
&&\psi_{2,n}(\vbr)=\sum_{l,m}d_{nlm}Y_{l,m}(\theta,\phi).
\ee
This transforms the HF equation to a matrix equation. The large matrix is splitting into several diagonal blocks for different values of the magnetic quantum number $m$. For a given $m$, the HF equation can be written as
\be\label{Eq:HFeq}
&&(T_{1,m}+V_{2,m})_{ll'}c_{nl'm}=E_n^{(1)} c_{nlm}, \nonumber \\
&&(T_{2,m}+V_{1,m})_{ll'}d_{nl'm}=E_n^{(2)} d_{nlm}
\ee
with the following matrix elements
\be
&&(T_{a,m})_{ll'}=\frac{l(l+1)m_1}{2m_a}\delta_{ll'},~ l=m,m+1,\cdots,l_{max}.\\
&&(V_{a,m})_{ll'}=g_{12}\int_{-1}^1dx \rho_a(x)N_{lm}P^m_l(x)N_{l'm}P^{m}_{l'}(x).
\label{Eq:Velement}
\ee
Here $x=\cos\theta$ and the normalization factor is $N_{lm}=\sqrt{\frac{(l-|m|)!}{(l+|m|)!}\frac{2l+1}{2}}$. In computing Eq.~\eqref{Eq:Velement}, we have used the definition of spherical harmonics as $Y_{lm}(\theta,\phi)=\frac{1}{\sqrt{2\pi}}N_{lm}P^m_l(\cos\theta)e^{im\phi}$, and the densities are assumed to respect azimuthal symmetry. By expanding the wavefunctions in terms of the Legendre polynomials, we have equivalently imposed the natural boundary condition at the north and south poles of the sphere to ensure the wavefunctions are smooth and have no singularities on the whole sphere.
The highest occupied angular momentum numbers are assumed to be $L_a$ with $a=1,2$. Then
\be\label{Eq:Na}
N_a=\sum_{l=0}^{L_a}(2l+1)=(L_a+1)^2.
\ee
Then we can take the highest energy level of the lighter fermions as the Fermi energy, which is given by $E_F/E_0=L_1(L_1+1)$.

Self-consistent solutions to Eq.~\eqref{Eq:HFeq} are obtained by implementing an iteration method similar to Ref.~\cite{parajuli2019mass}, which has also been implemented in Refs.~\cite{Chern2014dynamically,parajuli2023atomic,Fernando24}. We begin with a set of trial functions for $\rho_a$ and find the eigenfunctions and eigenvalues of Eq.~\eqref{Eq:HFeq}. Then a new set of $\rho_a$ can be assembled by Eq.~\eqref{Eq:rho}, which will be used in the next run. The iteration stops when the convergence condition $\int_{S^2}\sum_{a=1,2}|\rho_a^{new}-\rho_a^{old}| < \epsilon$ is met. We have checked that the convergent solution is in general insensitive to the initial trial density profiles. Interestingly, the iteration method is similar to that for solving the Bogoliubov-de Gennes equation for inhomogeneous Fermi superfluids~\cite{degennes-sc,BdG-book,Parajuli23,He_2024}, but there is no pairing gap in the repulsive Fermi gases.

Interestingly, the uniform-density profiles always satisfy the HF equations. Therefore, we have compared the total energies of the uniform solution and the phase-separation solution to find the energetically stable configuration.
Moreover, if multiple phase-separation solutions (like two-chunk or three-chunk structures) are obtained, we also compare their energies to pick the lowest-energy solution. 
For the uniform solution, the total energy is given by
\be
\frac{E_{uni}}{E_0}&=&\sum_{a=1}^{2}\sum_{l_a=0}^{L_a}\frac{m_1}{m_a}l_a(l_a+1)(2l_a+1)+
g_{12}N_1N_2.
\ee
Here $L_{1,2}$ are determined by the fermion numbers via Eq.~\eqref{Eq:Na}. For the phase-separation solution, the total energy is given by
\be
E_{sep}=\sum_{n=1}^{N_1}E_n^{(1)}+\sum_{n=1}^{N_2}E_n^{(2)}.
\label{E_sep}
\ee
Here $E_n^{(1)}$ and $E_n^{(2)}$ are the eigen-energies from Eq.~(\ref{eq-HF}). Numerically, we found that the phase-separation solution, if exists, always has lower energy than the corresponding uniform solution.

\begin{figure}
\centering
\includegraphics[width=\columnwidth]{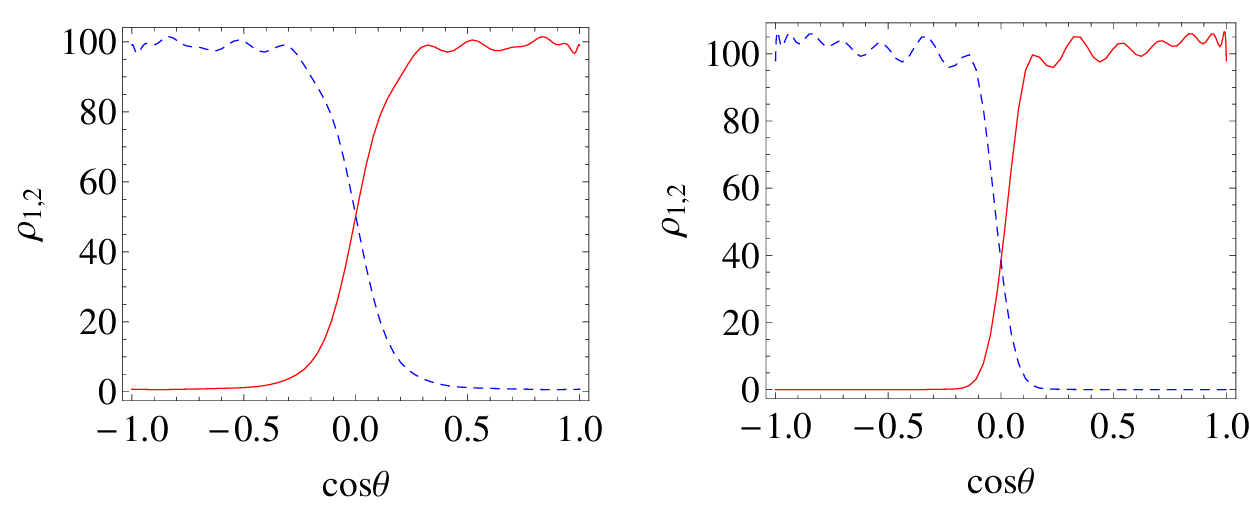}
\caption{The dimensionless densities $\rho_1=A\ep{n_1}$ (red solid line) and $\rho_2=A\ep{n_2}$ (blue dashed line) of equal-mass repulsive Fermi gases in a thin spherical shell as functions of $\cos\theta$ for $g_{12}=2$ (left) and $g_{12}=3$ (right). Here $A=4\pi R^2$ is the surface area. The particle numbers are $N_1=N_2=100$.
}
\label{dens}
\end{figure}

Next, we consider rotation of the atoms on the spherical shell along a fixed axis, which will be chosen as the $z$-direction, and find the structures of the Fermi mixtures. In the rotational frame, the Hamiltonian becomes~\cite{LandauMechanics}
\be
H'=H-\tilde{\omega}_z L_z,
\ee
which applies to quantum systems with suitable quantization. For a given magnetic quantum number $m$, the operator $L_z$ expanded by  spherical harmonics becomes a constant matrix. Therefore, the HF equation in the rotational frame is given by
\be\label{Eq:HF-r}
&&\Big[(T_{1,m}+V_{2,m})_{ll'}-\omega_z m\delta_{ll'}\Big]c_{nl'm}=E_n^{(1)} c_{nlm}, \nonumber \\
&&\Big[(T_{2,m}+V_{1,m})_{ll'}-\omega_z m\delta_{ll'}\Big]d_{nl'm}=E_n^{(2)} d_{nlm},
\ee
where the matrices $T_{1,2}$ and $V_{1,2}$ are the same as those without rotation, and the dimensionless angular velocity is $\omega_z=\hbar\tilde{\omega}_z/E_0$. Following a similar iteration method, the self-consistent solution can be obtained for the system with rotation.


\section{Results}\label{Sec:Results}
\subsection{Without rotation}
We first present the result of two-component fermions with equal masses $m_1=m_2$, which model Fermi gases of the same species but prepared in two different hyperfine states. We begin with the equal-population case where the fermion numbers are set to $N_1=N_2=100$ and the basis of angular-momentum states is capped at $l_{max}=20$. When $g_{12}$ is small, only uniform solutions can be found. However, phase separation becomes the stable solution as $g_{12}$ crosses a critical value. The densities assume azimuthal symmetry, thereby only their dependence on the polar angle $\theta$ is presented. In Fig. \ref{dens}, we plot the dimensionless density profiles of $\rho_1$ and $\rho_2$ as a function of $\cos\theta$ for $g_{12}=2$ (left panel) and $g_{12}=3$ (right panel). One can see that two-chunk separation with equal size of occupation occurs when the repulsion favors phase separation. As $g_{12}$ increases, the interface width between the two species shrinks to reduce the overlap, thereby helping minimize the interaction energy. Moreover, we have checked further increasing $l_{max}$ does not lead to observable changes. 

\begin{figure}
\centering
\includegraphics[width=\columnwidth]{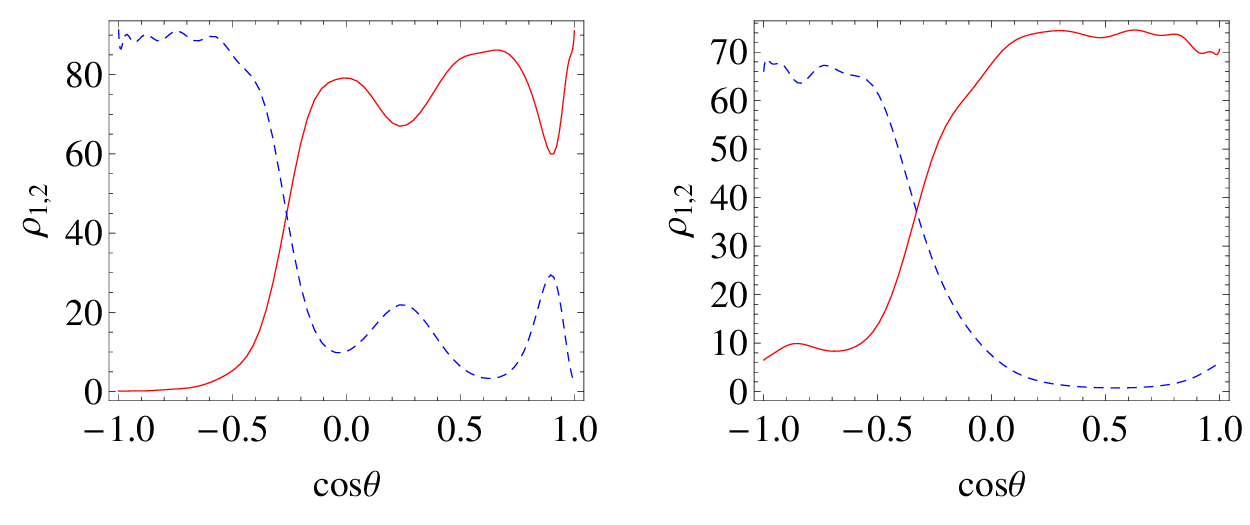}
\caption{The dimensionless densities $\rho_1$ (red solid line) and $\rho_2$ (blue dashed line) as functions of $\cos\theta$ of equal-mass repulsive Fermi gases with population imbalance in a thin spherical shell. Here $N_1=100$ and $g_{12}=2$ with $N_2=80$ (left) and $N_2=50$ (right).
}
\label{dens-N}
\end{figure}

For the equal-mass case, we also consider population imbalance of the two components by fixing $N_1$ and changing $N_2$. As shown in Fig.~\ref{dens-N}, population imbalance breaks the mirror symmetry along the equator because the majority component is expected to occupy a larger area.

\begin{figure}
\centering
\includegraphics[width=\columnwidth]{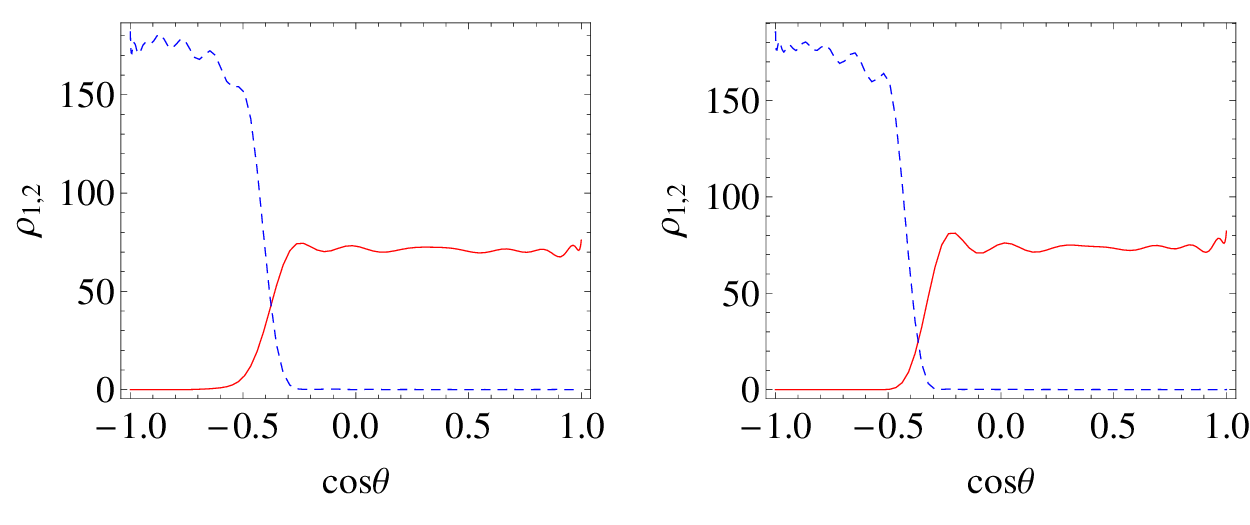}
\caption{The dimensionless densities $\rho_1$ (red solid line) and $\rho_2$ (blue dashed line) of mass-imbalanced repulsive Fermi gases on a spherical surface as functions of $\cos\theta$. Here $m_2/m_1=40/6$, $N_1=N_2=100$, and $g_{12}=1$ (left) and $g_{12}=3$ (right).}
\label{dens-m}
\end{figure}

Next, we introduce mass imbalance by considering two different species of fermionic atoms trapped in the same spherical shell. We consider two common fermionic species, $^6$Li and $^{40}$K, used in cold-atom experiments~\cite{Pethick-BEC,Ueda-book}. We focus on their dependence of the polar angle $\theta$ because the densities assume azimuthal symmetry. In Fig. \ref{dens-m}, we show the structures of mass-imbalanced cases with $m_2/m_1=40/6$ and equal populations on a spherical surface. One can see that the lighter fermions occupy a larger area than the heavier fermions do. This is because the mass appears in the denominator of the kinetic-energy term. Therefore, the lighter species with relatively higher kinetic energy is able to push the heavier species away and covers a larger area on the sphere.

\begin{figure}
\centering
\includegraphics[width=\columnwidth]{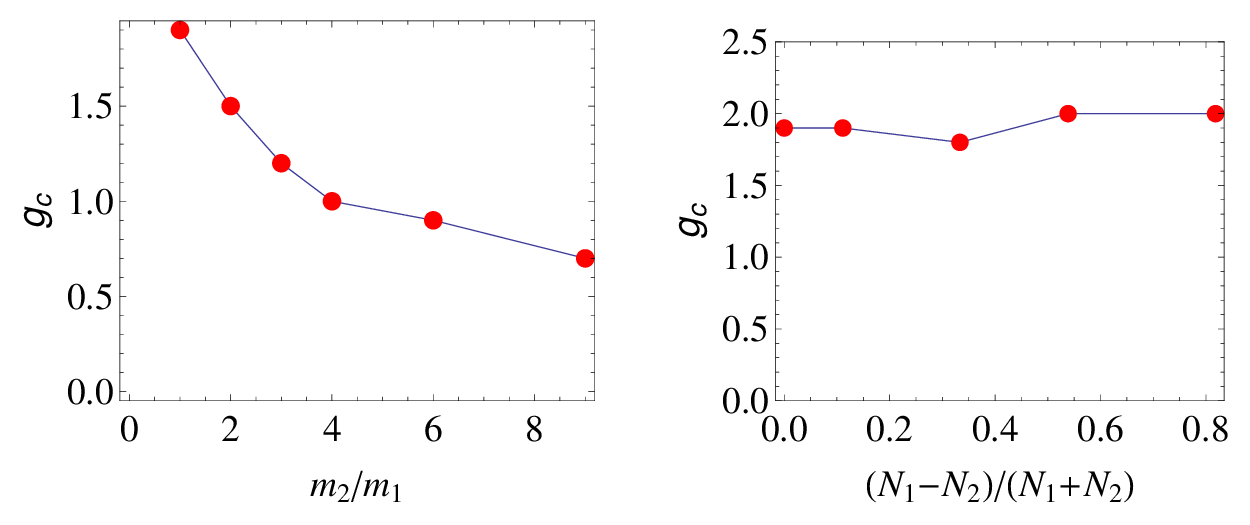}
\caption{Phase diagrams of two-component repulsive Fermi gases on a spherical surface. The left panel shows the critical $g_c$ as a function of $m_2/m_1$ with equal populations, and the right panel shows $g_c$ as a function of population imbalance with $m_1=m_2$. The system is a uniform mixture (exhibits phase separation) below (above) $g_c$.
}
\label{gc}
\end{figure}

There is a critical coupling constant $g_c$, above which the phase-separation solutions are favored. We found that $g_c$ decreases when the mass ratio $m_2/m_1$ increases. The left panel of Fig. \ref{gc} shows the phase diagram of mass-imbalanced Fermi gases equal-population in a spherical shell.
On the other hand, $g_c$ for the equal-mass case is quite insensitive to the presence of population imbalance. The right panel of Fig. \ref{gc} shows the phase diagram of the equal-mass Fermi gases with population imbalance, where $g_c$ is basically flat as the population imbalance increases. The contrast between the dependence of $g_c$ on mass imbalance and population imbalance suggests that tuning the kinetic energy by different masses is more efficient in inducing phase separation.

\subsection{Size-induced structural change}
The construction of the dimensionless coupling constant $g_{12}$ involves the two-body scattering length $a$ and the radius of the sphere $R$. While it is intuitive to vary the scattering length to tune the interaction and keep the size of the sphere fixed, one may as well fix $a$ and vary $R$ to control the interaction. Here we study how the density distributions of Fermi gases with repulsion depend on the radius of the sphere by varying $R$ with $a$ fixed.
We note that the dimensionless coupling $g_{12}\propto (R/a)^2$ according to Eq.~(\ref{eq-g12}). 
If $a$ is fixed and $R$ increases, the coupling $g_{12}$ in the dimensionless HF equation increases, and the system favors the phase-separation solution.

\begin{figure}
\centering
\includegraphics[width=0.6\columnwidth]{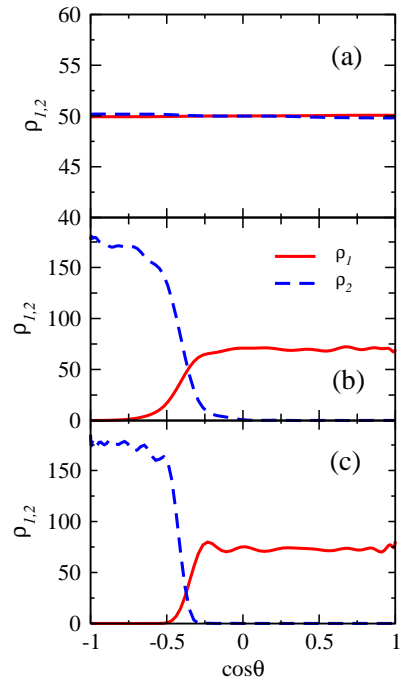}
\caption{The dimensionless densities $\rho_1$ (red solid lines) and $\rho_2$ (blue dashed lines) of repulsive Fermi gases on a spherical surface as functions of $\cos\theta$ with different values of the radius. (a) $R=0.5R_0$. (b) $R=0.63R_0$. (c) $R=R_0$. The mass ratio is $m_2/m_1=40/6$ and $g_{12}\approx 0.5, 0.8, 2.0$ from (a) to (c) due to the increasing radius. The particle numbers for all panels are $N_1=N_2=100$.
}
\label{fig-dens-R}
\end{figure}

Selected density profiles from different values of $R$ are shown in Fig. \ref{fig-dens-R} with a mass ratio $m_2/m_1=40/6$. For fixed scattering length $a$ and cutoff $\Lambda$, we take $R_0=1.32\Lambda^{1/2}a$ as a reference radius. Fig. \ref{fig-dens-R} (c) shows that when $R=R_0$, $g_{12}\approx2$ and the system has a two-chunk phase-separation solution as its lowest-energy configuration. For $R=0.5R_0$, the coupling $g_{12}\approx0.5$, and there is only a uniform solution as shown in panel (a). For $R=0.63R_0$ shown in panel (b), the coupling $g_{12}\approx0.8$ is very close to the critical value, and the phase-separation solution is about to turn into a uniform one if we reduce $R$ further. Similar structural changes are also found in equal-mass cases with or without population imbalance due to the same scaling of $g_{12}$ with respect to $R$.
We mention that Ref.~\cite{He22Sphere} shows size-induced crossover of the nature of the superfluid ground states of attractive two-component Fermi gases on a spherical surface instead of structural changes. 

\subsection{With rotation}
In the presence of rotation with constant angular velocity, the equal-mass case with equal population does not exhibit any qualitative difference from the corresponding case without rotation. The basic picture remains that uniform mixtures survive below $g_c$ and two-chunk separation emerges above $g_c$. This is because the two components with equal mass reacting to rotation in the same way, so they do not discern such a global effect. More specifically, the rotation makes the fermion density slightly higher around the equator than the two poles of the sphere. However, the density distributions of the two components of fermions are still identical. Therefore, the key factor to generate spatial separated solution is the strong repulsion between the fermions, which is basically the same as the situations without rotations. Similar situations also occur to population-imbalanced two-component Fermi gases with equal masses, so the solution with rotation is virtually the same as the corresponding solution without rotation despite the unequal density profiles of the two components due to population imbalance.

\begin{figure}
\centering
\includegraphics[width=\columnwidth]{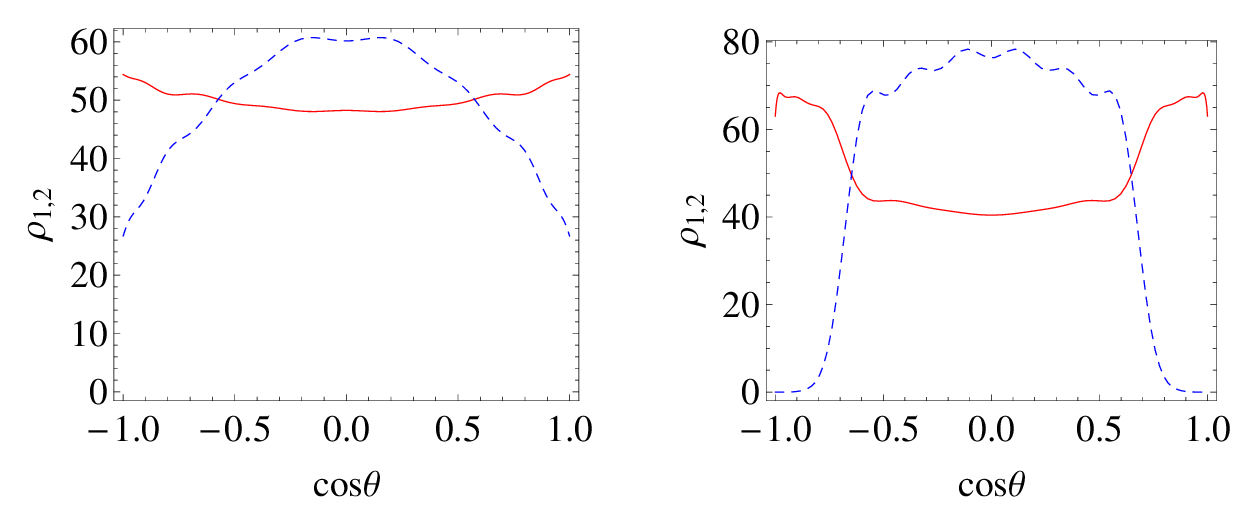}
\caption{The dimensionless densities $\rho_1$ (red solid lines) and $\rho_2$ (blue dashed lines) of mass-imbalanced repulsive Fermi gases in a thin spherical shell with rotation as functions of $\cos\theta$. Left panel:  $g_{12}=0.5$ and $\omega_z=2$. Right panel: $g_{12}=0.7$ and $\omega_z=4$. Here $m_2/m_1=40/6$ and $N_1=N_2=100$.}
\label{fig-om}
\end{figure}

In contrast, different structures are found for the mass-imbalanced case of $m_2/m_1=40/6$ with rotation in certain regime.
The density profiles of selected values of $\omega_z$ of two-species Fermi gases with $m_2/m_1=40/6$ and $g_{12} < g_c$ are shown in Fig. \ref{fig-om}. We find that instead of uniform density distributions, the heavier fermions occupy the region around the equator while the lighter fermions occupy the regions around the north and south poles when $\omega_c$ is above a critical value. The Fermi gas then has a sandwich (or three-chunk) structure, which is otherwise impossible without rotation or mass imbalance. When the repulsion is above $g_c$, however, the two-chunk phase separation solutions become stable, and there is no observable difference from the corresponding case without rotation in this regime. 

The emergence of the sandwich solution in the presence of rotation and mass-imbalance when $g_{12}<g_c$ may be understood from classical- or quantum-physics points of view. We begin with the classical picture. In the rotation frame, the extra term corresponding to an additional centrifugal potential energy $-I\omega_z^2$ with the moment of inertia $I$. If the heavier particles are located around the equator with the lighter particles pushed to the north and south poles, $I$ reaches a larger value than the homogeneous configuration, and the total energy in the rotation frame is lowered by forming the sandwich structure. 

From the quantum description, the extra term in the Hamiltonian in the rotation frame is $-\omega_zL_z$ determined by the angular momentum along the rotation axis. When the operator $L_z$ is applied to the basis of spherical harmonics, it gives integer eigenvalues corresponding to the magnetic quantum number $m=-l,\cdots l$. We note that the Legendre polynomial $P^m_l(x)$ typically has larger values around the two poles of the sphere for $m\approx 0$. On the other hand, if $|m|\lesssim l$, the larger values of $P^m_l(x)$ occur mostly around the equator. Therefore, if the wavefunctions of the heavier particles are mostly confined around the equator, the energy of the heaver particles can be lowered by the term $-\omega_zL_z$ while its kinetic-energy increase is smaller than that of the lighter species since the mass appears in the denominator. Moreover, two-chunk solutions do not optimize the configuration according to the above argument and were not found in our numerical calculations when $g_{12}<g_{c}$. Therefore, the sandwich solution is energetically favorable by rotation and mass-imbalance.

\begin{figure}
\centering
\includegraphics[width=0.8\columnwidth]{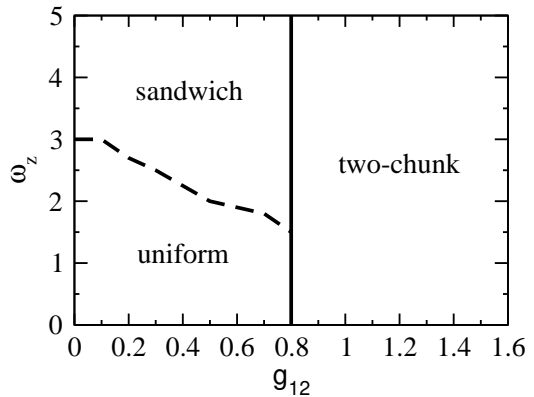}
\caption{Phase diagram on the $g_{12}$ - $\omega_z$ plane of repulsive two-species fermions with a mass ratio $m_2/m_1=40/6$ in a thin spherical shell with rotation. Here $N_1=N_2=100$. The vertical solid line indicates $g_c$.
}
\label{fig-g-om}
\end{figure}

With the mass ratio $m_2/m_1=40/6$, the phase diagram in the parameter space of $g_{12}$ and $\omega_z$ according to the density distributions are shown in Fig.~\ref{fig-g-om}. The two-chunk phase has configurations similar to those in Fig. \ref{dens-m} and dominates the regime when $g_{12}>g_c$ even in the presence of rotation. Here $g_c\approx0.8$ is about the same as the critical coupling without rotation. To understand why $g_c$ is insensitive to rotation, we may consider an extreme example with $g_{12}=0$. Since the extra term $-\omega_z L_z$ in the rotation frame acts like a Zeeman splitting term when a magnetic moment encounters a magnetic field, one expects that 
the eigenstates with large positive magnetic quantum numbers (large values of $m$) will be occupied to lower the energy of the ground state. Therefore, both species of particles will tend to accumulate around the equator to enlarge the number of high-$m$ states. To achieve a spatially separated solution when $g_{12}$ is large, the vital factor is only the repulsion between the two types of fermions since the rotation does not favor any component. Thus, $g_c$ is roughly the same in the rotation frame. 
Nevertheless, stable sandwich structures start to emerge with large $\omega_z$ when $g_{12}<g_c$ as shown in Fig.~\ref{fig-g-om}. In the sandwich phase, the structure is similar to those illustrated in Fig. \ref{fig-om}, which has been explained previously. Finally, the uniform phase with homogeneous mixtures survives in the regime with small $g_{12}$ and $\omega_z$.
 
We have a few remarks: When $g_{12}>g_c$ in the $m_2/m_1=40/6$ case, sandwich solutions may be found in the two-chunk regime but the two-chunk structures have slightly lower total energies. This observation suggests that the interface energy in the phase-separation solution is positive but relatively small, so the system minimizes the number of interfaces and has the two-chunk structure as the lower-energy solution. In contrast, there is no two-chunk solution
below the critical coupling $g_{12}$. However, stable sandwich solutions may emerge when $\omega_z$ is large enough in this regime. Therefore, the sandwich structure is made possible by rotation and mass imbalance. Meanwhile, the dashed line of Fig.~\ref{fig-g-om} only indicates where the sandwich structure starts to emerge. As shown in Fig.~\ref{fig-om}, the relative difference in densities may be small for the sandwich solution when $\omega_z$ is slightly above the dashed line of Fig.~\ref{fig-g-om}.

\section{Discussion}\label{Sec:Discussion}
The theory of spherical bubble traps has been developed earlier~\cite{PhysRevLett.86.1195,Lundblad2019} by considering mixing hyperfine states in a harmonic trap via radio-frequency excitation. Ref.~\cite{He22Sphere} generalized this idea with a sketch of possible spherical bubble traps for multi-component Fermi gases of the same species. It may be challenging to experimentally trap different species of atoms in the same spherical shell. Nevertheless, the rapid development of techniques for trapping ultracold atoms may resolve this issue in the future. Multi-component atomic gases in the spherical-shell geometry may also introduce interesting topological excitations~\cite{He2023Soliton} and bridge different areas of research.

There have been various ways for inducing rotation of ultracold atoms. For example, light-atom interactions may act as an utensil to stir the atomic cloud. This method has shown its versatility in the studies of rotating BEC in spherical harmonic traps~\cite{PhysRevLett.83.2498,PhysRevLett.84.806,doi:10.1126/science.1060182,PhysRevLett.88.010405,Zwierlein2005} and ring-shape geometries~\cite{PhysRevLett.99.260401,PhysRevLett.124.025301,PhysRevLett.128.150401,PhysRevX.12.041037}. We remark that while quantum vortices are expected to emerge in the superfluid ground state of attractive Fermi gases on a sphere~\cite{He23Vortex} similar to those on a 2D plane, there is no superfluid in the ground state of the repulsive Fermi gases studied here, which explains the absence of quantum vortices in this work. In contrast, inhomogeneous structures emerge in the repulsive case due to the competition between kinetic and interaction energies when the atoms are rotating. 

The healing length is a characteristic length determining how rapidly the density returns to the bulk value across the phase separation interface, which has been studied in Ref.~\cite{parajuli2023atomic} for boson-fermion mixtures. Using a similar argument based on the competition between the kinetic and interaction energies at the interface separating the two chunks, here we quantify the healing lengths of the repulsive Fermi gases. For component $j=1,2$ of the Fermi gas, the kinetic energy density due to the density distortion at the interface with a characteristic length $\xi_j$ is $\hbar^2 n_j/(2m_j \xi_j^2)$ while the interaction energy density is $gn_1 n_2 E_0$. The interface is where the two energies are comparable, thereby giving the estimation $\hbar^2/(2m_j \xi_j^2)\approx gn_{3-j} E_0$. Therefore, $\xi_{1,2}/R=\sqrt{m_1/[m_{1,2}g\rho_{2,1}/(4\pi)]}$. Taking the case shown in Fig.~\ref{dens} (a) as an example, the healing length $\xi_1$ with $g=2$ and $N_2=100$ is $\xi_1/R= \sqrt{1/[2 \rho_2/(4\pi)]}\approx 0.25$, which spans a small arc of $4\%$ of a great circle on the sphere. Moreover, increasing $g$ while fixing $n_2$ leads to a smaller $\xi_1$, as the expression implies. One can see that this is indeed the case, as the density profiles become sharper at the phase-separation interface as $g$ increases from Fig.~\ref{dens} (a) to (b). The presence of population- or mass- imbalance only introduces quantitative corrections while the general behavior of narrower interface width as $g$ or $n_{1,2}$ increases still holds in Figs.~\ref{dens-N} and \ref{dens-m}. We caution, however, the sandwich structures due to rotation shown in Fig.~\ref{fig-om} do not deplete the densities completely across different regions, so the concept of healing length may not fully apply.

We also comment on some assumptions behind our work. Although we present the ground-state properties of repulsive Fermi gases on a spherical surface at zero temperature, it is possible to generalize to finite temperatures by including the Fermi-Dirac distributions in the thermodynamic quantities and adding the entropic contribution to the free energy. Since finite temperatures amplify the kinetic-energy effects in general, homogeneous mixtures will be favored and $g_c$ is expected to increase with temperature. Moreover, the interface between different components in phase separation will also be broadened as temperature increases. Nonetheless, the general mechanism of competing kinetic and interaction energies is still at play in determining the structures of the mixtures. We also assume infinitely thin shells in our study, which correspond to a strong confinement along the radial direction. If the thickness of the shell is not negligible, one has to include the radial equation in the HF equation. By increasing the shell thickness, a previous study on BEC suggests the results will eventually resemble those in spherical harmonic potentials~\cite{PhysRevA.98.013609}. For repulsive Fermi gases in spherical harmonic potentials, possible structures have been studied in  Refs.~\cite{PhysRevLett.110.165302,Pecak16,Bellotti2017}.

\section{Conclusion}\label{Sec:Conclusion}
In summary, we have found the ground states of two-component repulsive Fermi gases with and without population or mass imbalance on a thin spherical shell. Homogeneous or two-chunk structures are stable when the repulsion is below or above the critical value of repulsion, respectively. When the atoms on the sphere are rotating, mass-imbalance is shown to induce sandwich structures with the heavier (lighter) species concentrated around the equator (north and south poles) below the critical repulsion due to maximization of the total angular momentum along the rotation axis. The sandwich structure on a sphere is different from that in a box because the latter is due to the distortion of the wavefunctions at the hard walls. Those interesting phenomena of atomic mixtures in a spherical shell will offer inspirations for future research on interacting quantum systems in curved geometries. 

\begin{acknowledgments} 
Y. H. was supported by the NNSF of China (No. 11874272). C. C. C. was partly supported by the NSF (No. PHY-2310656).
\end{acknowledgments}

%

\end{document}